\documentclass[useAMS,usenatbib]{mn2e}

\usepackage{graphicx}
\usepackage{amsmath}
\usepackage{xspace}
\usepackage{dsfont}

\renewcommand{\btheta}{\{\theta_i\}}
\newcommand{\bx}{\{x_i\}}

\title[Hierarchical Reverberation Mapping]
{Hierarchical Reverberation Mapping}
    
\author[Brewer and Elliott]{%
  Brendon~J.~Brewer$^{1}$\thanks{bj.brewer@auckland.ac.nz},
  Tom M. Elliott$^{1}$
  \medskip\\
  $^1$Department of Statistics, The University of Auckland, Private Bag 92019, Auckland 1142, New Zealand}

\begin{document}
             
\date{To be submitted to MNRAS Letters}
             
\maketitle

\label{firstpage}

%%%%%%%%%%%%%%%%%%%%%%%%%%%%%%%%%%%%%%%%%%%%%%%%%%%%%%%%%%%%%%%%%%%%%%%%%%%%%%

\begin{abstract}
Reverberation mapping (RM) is an important technique in studies of active
galactic nuclei (AGN). The key idea of RM is to measure the time lag $\tau$
between variations in the continuum emission from the accretion disc
and subsequent response of the broad line region (BLR). The measurement of
$\tau$ is typically used to estimate the physical size of the BLR and is
combined with other measurements to estimate the black hole mass $M_{\rm BH}$.
A major difficulty with RM campaigns is the large amount of data needed to
measure $\tau$. Recently, \citet{2012MNRAS.427.2701F} introduced a new approach
to RM where the BLR light curve is sparsely sampled, but this is counteracted
by observing a large sample of AGN, rather than a single system.
The results are combined to infer properties of the sample of
AGN. In this letter we implement this method using a hierarchical
Bayesian model and contrast this with the results from the previous stacked
cross-correlation technique. We find that our inferences are more precise and
allow for more straightforward interpretation than the stacked cross-correlation
results.
\end{abstract}


%%%%%%%%%%%%%%%%%%%%%%%%%%%%%%%%%%%%%%%%%%%%%%%%%%%%%%%%%%%%%%%%%%%%%%%%%%%%%%

\section{Introduction}
Reverberation mapping (RM) is a key technique for the study
of active galactic nuclei (AGN). The technique is based on the temporal
fluctuations of the central continuum source, and the subsequent response
of the broad line region (BLR) emission. The time delay between the continuum
and the broad line fluctuations provides an estimate of the size of the BLR,
and can also be used to estimate the black hole mass \citep{peterson}.

RM is a observationally intensive, requiring observations of an AGN
over a period of
a few tens of days \citep{2013ApJ...769..128B}. As a result, many authors have
studied the data analysis techniques involved in RM, and substantial
advances have been made in recent years.
The data analysis methods introduced range from those that attempt to infer the
transfer function \citep[the distribution of lags in a single object,
e.g.][]{1995ApJ...440..166K, 2011ApJ...735...80Z},
the velocity-resolved transfer function
\citep{2010ApJ...720L..46B}, or the physical structure of the BLR itself
\citep{pancoast, arp151, pancoast2, 2013arXiv1310.3907L}.

There are many subtleties and challenges involved in reverberation mapping,
that we will ignore
for the purposes of this letter. We note two of them here for completeness. Firstly,
the mean lag $\bar{\tau} = \int \tau\Psi(\tau)\, d\tau$, where $\Psi(\tau)$
is the normalised transfer function, is not equal to $c$ times the mean radius of the BLR
matter distribution, $\bar{r} = \int \sqrt{x^2+y^2+z^2}\rho(x, y, z)\, d^3 \mathbf{x}$.
Secondly, the mean lag $\bar{\tau}$ is not equivalent to the peak of the
cross-correlation function, except in the case of very narrow
transfer functions. Direct physical modelling of the BLR resolves these
issues \citep{pancoast, arp151, 2013arXiv1310.3907L}.

Recently, \citet{2012MNRAS.427.2701F, 2013MNRAS.434L..16F} introduced an
innovative approach to reverberation mapping where the results from multiple
AGN can be combined to yield inferences about the entire sample of AGN,
despite the fact that the constraints on any individual AGN are poor. Rather
than accurately measuring $\tau$ in a single object, it is possible to roughly
measure $\tau$ for a large number of objects (for example, by only measuring
the BLR emission line flux at two epochs), and to infer properties about
the distribution of $\tau$ values in the sample of objects (and hence in
a broader population, if the sample can be considered representative).
The data analysis approach of \citet{2012MNRAS.427.2701F} was based on an
extension of the traditional cross-correlation method to $N$ separate objects.
By ``stacking'' the cross-correlation functions (CCFs) from the $N$ objects in
the sample, a peak appears which indicates a typical lag value in the sample
of AGN. However, this peak is difficult to interpret. Specifically, the width
of the peak may be influenced either by uncertainty (due to the sparse data)
or due to a wide range of lags being present in the sample (i.e. the sample of
AGN is diverse).

In this letter, we revisit the idea of ``stacked'' reverberation mapping (i.e.
using a sample of AGN with sparse data to infer something about the sample)
from a Bayesian perspective. We develop a Bayesian hierarchical model
\citep{loredo, kelly, extreme_deconvolution, 2012AJ....143...90S, 2012arXiv1208.3036L, 2013arXiv1310.5177B, 2013AJ....146....7B}.
which allows us to calculate the posterior distribution for hyperparameters
which describe the typical lag, and the diversity of the lags.

In Section~\ref{sec:multiple} we discuss the general principles needed
to combine information from multiple poorly-measured objects.
In Section~\ref{sec:single_object} we discuss the methods used to infer the lag
$\tau$ using RM data for a single AGN. We then demonstrate the technique
on simulated data in Section~\ref{sec:data}, and conclude in
Section~\ref{sec:conclusions}.

\vspace{-0.5cm}
\section{Combining Inferences About Multiple Objects}\label{sec:multiple}
Consider a sample of $N$ objects, each of which has parameters $\theta_i$.
If the $i$th object is analysed on its own, the inference about 
its parameters $\theta_i$ is described by a posterior distribution
\begin{eqnarray}
p(\theta_i | x_i) \propto \pi(\theta_i)p(x_i | \theta_i)\label{eq:individual}
\end{eqnarray}
where $\pi(\theta_i)$ is the prior distribution and $p(x_i | \theta_i)$ is the
likelihood function. If $N$ objects are analysed separately, inferences can
be made about the diversity of $\theta_i$ across the sample. However, these
inferences may be incorrect due to the implicit assumption that the prior for
all of the $\btheta$ is independent \citep{2013arXiv1310.5177B}.
A hierarchical model may be used to overcome this problem.

In a hierarchical model, the prior for the parameters $\btheta$ is created
by introducing hyperparameters $\alpha$ such that the joint prior for $\alpha$
and the $\btheta$ is:
\begin{eqnarray}
p(\alpha, \btheta) &=& p(\alpha)\prod_{i=1}^N p(\theta_i | \alpha)
\end{eqnarray}
The implied prior on the $\btheta$ parameters is then
\begin{eqnarray}
p(\btheta) &=& \int p(\alpha)\prod_{i=1}^N p(\theta_i | \alpha) \, d\alpha
\end{eqnarray}
which may imply that the $\theta$ values are ``clustered'' around some
typical value.

The posterior distribution for
$\btheta$ and the hyperparameters $\alpha$
given the data 
$\{x_1, ..., x_N\}$, is then
\begin{eqnarray}
p(\alpha, \btheta | \bx) &\propto&
p(\alpha)p(\btheta|\alpha)p(\bx | \btheta, \alpha)\\
&=& p(\alpha)\prod_{i=1}^N f(\theta_i|\alpha)p(x_i | \theta_i)
\end{eqnarray}

It is possible to consider the data for all $N$ objects as one big data set
and infer the hyperparameters $\alpha$ and the parameters of all objects $\btheta$
simultaneously. However, this can be
computationally prohibitive. In some circumstances it is tractable to analyse
each object's data separately (using a common prior $\pi(\theta_i)$) and
then post-process the results, reconstructing what the results of the
hierarchical model would have been. In this letter we use this latter approach.

The marginal posterior distribution for the hyperparameters $\alpha$ is
\begin{eqnarray}
p(\alpha | \bx) &=&
\int p(\alpha, \btheta|\bx) \, d\btheta\\
&\propto& p(\alpha)\int \prod_{i=1}^N f(\theta_i|\alpha)p(x_i | \theta_i) d^N\theta\\
&\propto& p(\alpha) \prod_{i=1}^N \int f(\theta_i|\alpha)p(x_i | \theta_i) d\theta_i\\
&\propto& p(\alpha) \prod_{i=1}^N \int \frac{f(\theta_i|\alpha)}{\pi(\theta)}p(x_i | \theta_i) \pi(\theta_i)d\theta_i\\
&\propto& p(\alpha) \prod_{i=1}^N \mathds{E}\left[\frac{f(\theta_i|\alpha)}{\pi(\theta)}\right]\label{eq:combine}
\end{eqnarray}
where the expectation is taken with respect to the individual object posterior
of Equation~\ref{eq:individual}, and thus can be estimated using posterior
samples from the posterior distributions for the individual objects.
This result enables us to reconstruct the
posterior distribution for the hyperparameters even though the individual object
inferences were made without the hierarchical structure in the prior. This is
essentially an importance sampling approximation to the
full hierarchical model. One drawback of this approach is that it implies an
independent (non-hierarchical) prior on any per-object nuisance parameters,
which can sometimes adversely affect the results \citep[e.g.][]{2013arXiv1310.5177B}.

In Section~\ref{sec:single_object} we define a simple model for inferring the
lag of a single AGN from sparse RM data. We will use Markov Chain Monte Carlo
(MCMC) to produce posterior samples for the lags of the $N$ objects, which can
then be combined using Equation~\ref{eq:combine} to yield the posterior
distribution for some hyperparameters describing the distribution of
lags in the sample.

\vspace{-0.5cm}
\section{The Single Object Model}\label{sec:single_object}
If the continuum light curve of an AGN
is described by a function $y(t)$, then the line
light curve $l(t)$ is given by
\begin{eqnarray}
l(t) &=& A \int_\tau \Psi(\tau)\left[y(t - \tau) + C\right] \, d\tau
\end{eqnarray}
where $\Psi(\tau)$ is the transfer function (assumed to be normalised),
and $A$ and $C$ are response
coefficients. The idea of reverberation mapping is to use noisy measurements
of $y(t)$ and $l(t)$ to infer the transfer function $\Psi(\tau)$ or a summary
of it such as the mean lag $\bar{\tau} = \int \tau\Psi(\tau) \, d\tau$.
Throughout this letter we consider $y(t)$ and $l(t)$ in flux
units, as opposed to magnitudes, and consider $\bar{\tau}$ as
the definition of ``the lag'' of an AGN.

The posterior distribution for the lag $\bar{\tau}$ of a single object
$i$ may be obtained
by fitting the following model. We assume, for simplicity, that
the transfer function is uniform
between limits $a$ and $b$, where $b > a$:
\begin{eqnarray}
\Psi(\tau) &=& \left\{
\begin{array}{lr}
\frac{1}{b-a}, & \tau \in [a,b]\\
0, & \textnormal{otherwise}
\end{array}\right.
\end{eqnarray}
and our goal is to measure the mean lag:
\begin{eqnarray}
\bar{\tau} &=& \int \tau \Psi(\tau) \, d\tau\\
&=& \frac{1}{2}(b-a).
\end{eqnarray}
Note that inferring $\bar{\tau}$ from the data requires that we
marginalise over an infinite number of nuisance parameters describing the
behaviour of $y(t)$ at unobserved times \citep{pancoast}.
The prior for the underlying time variation of the continuum emission is
a continuous autoregressive process of order 1, or a CAR(1) model. These models
have been studied extensively for AGN variability
\citep[e.g.][]{2009ApJ...698..895K, 2011ApJ...735...80Z, 2013ApJ...765..106Z}.
This marginalisation can be done
either analytically or inside MCMC. We used the latter approach for simplicity.
We discretised
time using ten time bins per day, and have a discrete continuum light curve
$\mathbf{y} = \{y_1, ..., y_n\}$, included in the model as a set of unknown
parameters. The prior for these parameters is:
\begin{eqnarray}
p(y_i | y_{i-1},  m, k, \beta) \sim \mathcal{N}
\left(m + k\left(y_{i-1} - m\right), \beta^2\right)
\end{eqnarray}
for $i \geq 2$. This is the discrete AR(1) model from time series theory.
$m$ describes the mean level of the continuum light curve, $\beta$ controls
the size of the short-term fluctuations, and $k$ controls the correlation
timescale (the timescale itself is given by $L = -1/\ln(k)$).

For the likelihood (or sampling distribution, really the prior for the data
given the parameters) we made the conventional assumption of ``Gaussian noise'':
\begin{eqnarray}
Y_i &\sim& \mathcal{N}\left(y(t_{y_i}), \sigma_{y_i}^2\right)\\
L_i &\sim& \mathcal{N}\left(l(t_{l_i}), \sigma_{l_i}^2\right)
\end{eqnarray}

We used vague priors for the parameters of the single-object model. In practice
these could be made substantially narrower.
We chose a log-uniform prior for $L$ (the AGN variability timescale)
between $0.1$ and $10^5$ days, a log-uniform
prior for $\beta$ (the size of the short-term fluctuations)
between $10^{-3}$ and $10^3$ flux units, a Cauchy(0, 100)
prior for $m$ (the mean flux level of the AGN), a log-uniform prior for $b$
(the upper limit of the transfer function) between $10^{-4}$ and $10^2$ days,
a uniform prior for $a$ (the lower limit of the transfer function)
between 0 and $b$, a log-uniform prior between
$10^{-3}$ and $10^3$ for the response coefficient $A$ and a Cauchy(0, 100) prior
for the second response parameter $C$.

To implement the MCMC for a single object,
we implemented our model in the STAN sampler \citep{nuts}
and Diffusive Nested Sampling \citep{dnest}. Due to the large number of
parameters, multiple modes, and strong correlations in the posterior distribution,
we found that Diffusive Nested Sampling was more effective than STAN.
Note that our single-object model is already an improvement over standard
cross-correlation techniques and is similar to the approach used by
\citet{2011ApJ...735...80Z}.

\vspace{-0.5cm}
\section{Demonstration on Simulated Data}\label{sec:data}
To test our hierarchical model, and compare it to the stacked cross-correlation
function, we simulated data from a sample of 100 AGN. We assumed the
continuum flux for each object fluctuates around a mean value $m = 50$ in
arbitrary units. The variability timescales $L$ were simulated from a very
broad log-uniform distribution (i.e. a uniform distribution for $\ln(L)$)
with a lower limit of 10 days and an upper limit of $1,000$ days. The
variability amplitudes $\beta$ were also simulated from a log-uniform
distribution with lower limit of 0.2 flux units and an upper limit of 1 flux
unit. This implies some objects are more variable than others, as is typical
in reverberation mapping.

Each object in the sample had its own transfer function parameters. The upper
limit $b$ of the transfer function was simulated using a log-normal distribution
with a median value of 10 days, and the standard deviation of $\ln(b)$ was
0.3. Given $b$, $a$ was simulated from a uniform distribution between $0$ and
$b$. The resulting distribution of lags $\bar{\tau} = (a+b)/2$ is very well
approximated by a lognormal distribution with a median of 7.4 days and width
(quantified by the standard deviation of $\log_{10}(\bar{\tau})$) of 0.157
(these two quantities are the ones we will infer using the hierarchical model).
The response amplitudes $A$ and $C$ were set to 0.5 and 0 respectively, for
all objects. Note that the distribution of the parameters of the 100 objects is
not the same as the priors used when analysing them with the single-object model.

\begin{figure}
\begin{center}
\includegraphics[scale=0.45]{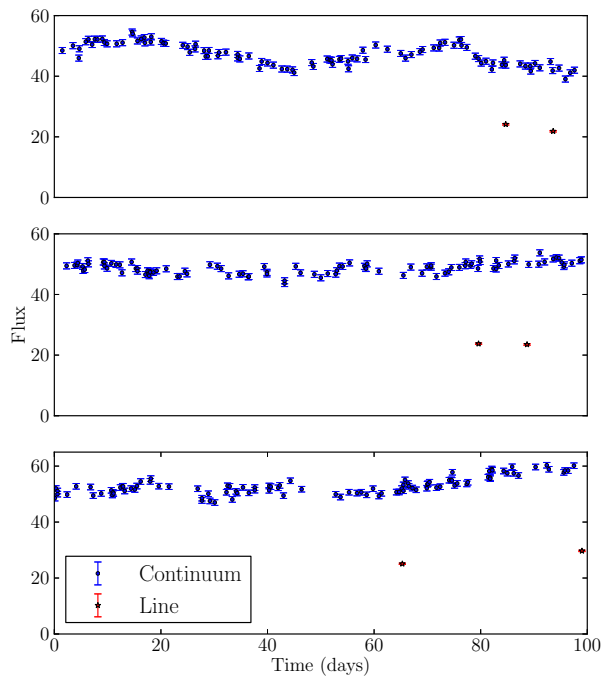}
\caption{Three simulated data sets, out of a total of 100. It is difficult to
infer the lag $\bar{\tau}$ for any of these objects (although there is some
information available). See Figure~\ref{fig:example_posteriors} for the
$\bar{\tau}$ posterior distributions for these three data sets.
\label{fig:data}}
\end{center}
\end{figure}

The measured data for each of the 100 AGN was
continuum flux measurements, once per day, for 100 days. The standard deviation
of the measurement noise is 1 flux unit (i.e. 2\%). The line data is measured on
only two days, with times selected from a uniform distribution between $t=50$
and $t=100$ days (i.e. corresponding to the latter half of the continuum data).
The flux of the line data is typically about half that of the continuum data
(since $A=0.5$ for all objects), and the
measurement noise for the line data is 0.25 flux units, or about 1\%.
See Figure~\ref{fig:data} for three illustrative data sets from the sample.

\begin{figure}
\begin{center}
\includegraphics[scale=0.35]{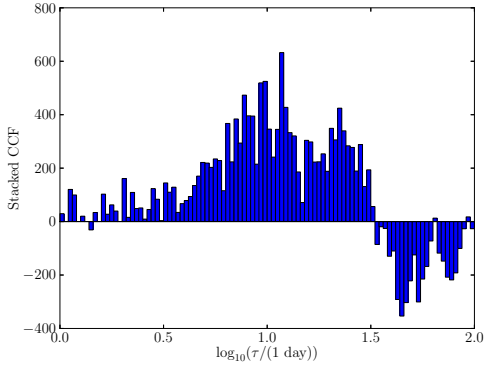}
\caption{The stacked cross-correlation function of the simulated data.
A clear peak is seen around $\tau \approx 10$ days, but the stacked CCF has a
large width. It is difficult to interpret the cause of this width, and to
determine whether the width is due to uncertainty or diversity in the sample
of AGN. These difficulties motivate the hierarchical Bayesian model.
\label{fig:ccf}}
\end{center}
\end{figure}

We computed the stacked cross-correlation function for the 100 data sets.
The cross-correlation function for a single object is defined as
in \citet{2012MNRAS.427.2701F}. For a lag bin between $\tau$ and $\tau + \delta$
the amount of cross correlation is calculated by looping over pairs of points
(one selected from the continuum data set, and another from the line data set),
and accumulating mass when the lag between the two points is between $\tau$
and $\tau + \delta$.
\begin{align}
& X(\tau; \delta) = \nonumber \\
& \frac{\sum_{i=1}^{N_y}\sum_{j=1}^{N_l}
\left(y_i - \bar{y}\right)
\left(l_j - \bar{l}\right)
\mathds{1}\left((t_{l_j} - t_{y_i}) \in [\tau, \tau + \delta]\right)}
{\sum_{i=1}^{N_y}\sum_{j=1}^{N_l}
\mathds{1}\left((t_{l_j} - t_{y_i}) \in [\tau, \tau + \delta]\right)}
\end{align}
The denominator is the number of pairs of points that fall within the lag range
being considered, $[\tau, \tau + \delta]$, denoted $n_{\rm pair}$ by
\citet{2012MNRAS.427.2701F}.
The stacked cross-correlation function for $N$ objects is sum of the CCFs
of the individual objects. In Figure~\ref{fig:ccf}, we show the stacked CCF
(on a log scale) from our 100 simulated AGN data sets.
There is a clear peak around 10 days, and the full width at half maximum,
while not strictly well defined, is roughly 0.25 dex.

As the data does not contain measurements at all possible time lags, the stacked
CCF will always require some smoothing. Here, the smoothing is supplied by the
choice of a bin width $\delta$. The stacked CCF has a substantial width. One
weakness of the stacked CCF method is that it is not clear whether this width
is caused by real diversity in the sample of AGN, or whether it is simply caused
by the difficulty in measuring the lag $\bar{\tau}$ of an object with such
sparse data. By contrast, the hierarchical Bayesian approach clearly separates
these two concepts.

\begin{figure}
\begin{center}
\includegraphics[scale=0.45]{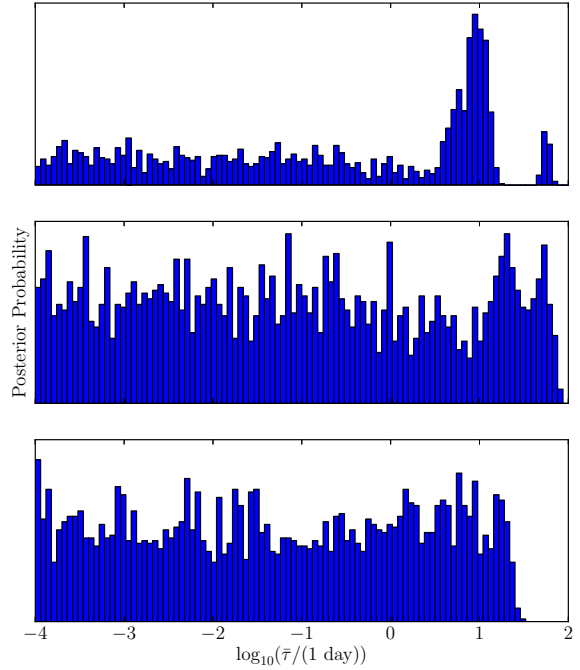}
\caption{Samples from the posterior distribution for $\log_{10}(\bar{\tau})$
for three objects whose data sets are shown in Figure~\ref{fig:data}.
These posteriors were obtained by analysing each object separately with the
single-object model. For
the first object, there is a moderate ``detection'' of a lag at about 10 days
(although there is a small second mode above 100 days, which can be understood
by inspecting the data in Figure~\ref{fig:data}).
The second object's lag was not well constrained,
and the third object's data only provided an upper limit on its lag.
\label{fig:example_posteriors}}
\end{center}
\end{figure}

To implement the hierarchical Bayesian approach on the simulated data, we first
used the single-object model on each object. The results for three objects
(the same three objects shown in Figure~\ref{fig:data}) are plotted in
Figure~\ref{fig:example_posteriors}. Due to the sparse data, there is a large
amount of uncertainty about the value of $\bar{\tau}$ for each system. Some
objects do not allow any real measurement of $\bar{\tau}$, but some do, as a
result of fortuitous variability of the continuum.

Finally, we combined the results from the single-object model into an inference
about hyperparameters $\mu$ and $\sigma$ describing the central value and
the diversity of $\bar{\tau}$ respectively, using the result of
Equation~\ref{eq:individual}. This requires us to know the prior $\pi(\bar{\tau})$
implied by the simple-object model, since we did not assign the prior directly
to $\bar{\tau}$ but indirectly through $a$ and $b$.
The prior for $\bar{\tau}$ in the single-object model is
very well approximated by a log-uniform distribution.

Since lags $\bar{\tau}$ are positive, 
our assumption for the prior on $\bar{\tau}$ given
the hyperparameters is a lognormal distribution with median $\mu$ and
width (standard deviation of $\log_{10}(\bar{\tau})$) $\sigma$:
\begin{eqnarray}
\log_{10}(\bar{\tau}) \sim \mathcal{N}(\log_{10}(\mu), \sigma^2)
\end{eqnarray}

\begin{figure}
\begin{center}
\includegraphics[scale=0.45]{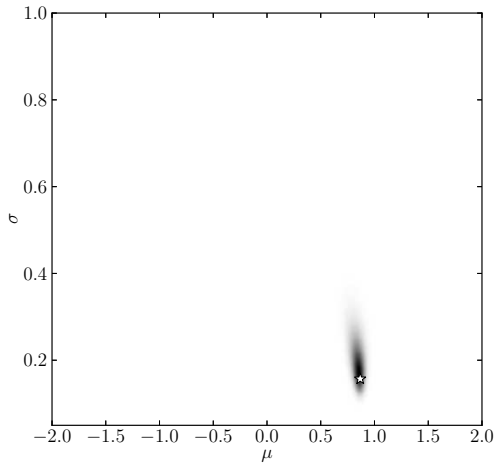}
\caption{The joint posterior distribution for $\mu$ and $\sigma$, hyperparameters
describing the distribution of $\bar{\tau}$ values in the sample. The prior
distribution was uniform over the area shown.
The true solution $(\mu = 0.867, \sigma = 0.157)$
is indicated by the white star symbol. $\mu = 0.867$ corresponds to a typical
lag of 7.4 days.\label{fig:posterior}}
\end{center}
\end{figure}

The joint posterior distribution for $\mu$ and
$\sigma$, given the data from $N=100$ AGN,
is shown in Figure~\ref{fig:posterior}. The prior for $\mu$ and $\sigma$
was uniform inside the rectangle shown, and the posterior distribution covers
a much smaller area, indicating that the data contained a lot of information
about the hyperparameters. The true values used to generate the simulated data
are denoted by the star symbol in Figure~\ref{fig:posterior},
and is typical of the posterior distribution, as it should be.

\begin{figure}
\begin{center}
\includegraphics[scale=0.3]{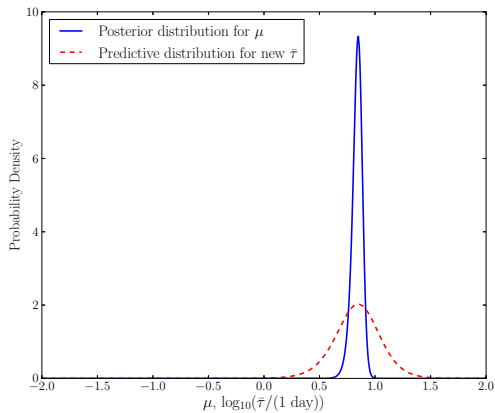}
\caption{The marginal posterior distribution for $\mu$, the 
hyperparameter describing the center of the distribution for
$\log_{10}\left(\bar{\tau}\right)$, is plotted as the solid line.
\label{fig:posterior2}}
\end{center}
\end{figure}

The marginal posterior distribution for $\mu$ is shown in Figure~\ref{fig:posterior2}
along with the posterior predictive distribution for the ``next'' $\bar{\tau}$ value. The inference about $\mu$ can be summarised as
$\mu = 0.838 \pm 0.047$ (consistent with the known input value of
$0.867 \approx \log_{10}(7.4)$),
and the prediction about $\log_{10}(\bar{\tau})$
may be summarised as $\log_{10}(\bar{\tau}) = 0.838 \pm 0.215$. The Bayesian
approach has clearly separated the uncertainty about $\mu$ from the diversity of the
sample, described by $\sigma$.

Note that our method for combining individual object results into an inference
about the population (Section~\ref{sec:multiple}) assumes that the hierarchical
prior applies to only a single parameter (in our case $\bar{\tau}$). If we
incorporate additional information by using a hierarchical prior for the other
model parameters, allowing us to detect (for example) that all of the response
amplitudes $A$ were equal to 0.5, we may be able to make more accurate
inferences. However, this would increase the complexity of the method and the
computational expense.

\vspace{-0.5cm}
\section{Conclusions}\label{sec:conclusions}
Reverberation mapping campaigns are observationally intensive, yet potentially
very informative. This has led to
research into RM data analysis techniques, as well as less costly observing
strategies. The recent approach of \citet{2012MNRAS.427.2701F} demonstrated
that it is possible to combine information from many sparsely-measured AGN
and infer properties of the sample rather than insisting on accurate results
about a particular object.

In this letter we extended the approach of \citet{2012MNRAS.427.2701F} by
implementing a Bayesian hierarchical model for the data analysis. The main
advantage of this approach is a clear interpretation of the output, which is
a posterior distribution for hyperparameters describing the sample of AGN.

We demonstrated this approach on a simulated data set consisting of 100 AGN
with well-measured continuum light curves but sparsely measured (two epochs per
AGN) broad line light curves. For the simulated data set we were able to obtain
an inference on $\mu$, a hyperparameter describing the typical value of the lag,
to within 0.047 dex (1-$\sigma$ uncertainty), whereas the stacked cross-correlation
function had a FWHM of $\sim$ 0.25 dex. This is possible because a subset of
the objects in the sample will have a measureable lag. We hope this approach
will facilitate informative studies of samples of AGN.

\vspace{-0.5cm}
\section*{Acknowledgements}
It is a pleasure to thank Tommaso Treu, Anna Pancoast (UCSB), and
Brandon Kelly (UCSB) for many
useful conversations about reverberation mapping. The STAN ({\tt mc-stan.org})
team provided helpful advice about the use of their software. BJB is partially
supported by the Marsden Fund (Royal Society of New Zealand). We are grateful
to the referee, Stephen Fine, for his helpful comments.

\end{document}